\newcommand{\beq}{\begin{equation}}
\newcommand{\eeq}{\end{equation}}
\newcommand{\beqns}{\begin{equation}}
\newcommand{\eeqns}{\end{equation}}
\newcommand{\beqar}{\begin{eqnarray}}
\newcommand{\bs}{\begin{eqnarray*}}
\newcommand{\eeqar}{\end{eqnarray}}
\newcommand{\es}{\end{eqnarray*}}
\newcommand{\beqml}{\begin{mathletters}}
\newcommand{\eeqml}{\end{mathletters}}
\newcommand{\clust}{\cal C}

\newcommand{\x}{\mbox{\boldmath $x$}}
\newcommand{\0}{\mbox{\boldmath $x$}_0}
\newcommand{\y}{\mbox{\boldmath $y$}}

\newcommand{\evm}{\hat{\mbox{\boldmath $e$}}_{\mu}}

\newcommand{\Ptilde}{\tilde{P}}

\newfont{\fancy}{msbm10 scaled\magstep1}
\documentstyle[aps,preprint,eqsecnum]{revtex}
\begin{document}
\draft
\preprint{PURD-TH-93-10}
\title{Eigenspectrum and localization for diffusion with traps}
\author{Achille Giacometti and Hisao Nakanishi}
\address{ Department of Physics, Purdue University,
West Lafayette, IN 47907 }
\date{\today}
\maketitle
\begin{abstract}
We investigate the {\em survival-return} probability distribution and
the eigenspectrum for the transition probability matrix,
for diffusion in the presence of perfectly absorbing traps distributed
with critical disorder in two and three dimensions.
The density of states is found to have a Lifshitz tail in the low frequency
limit, consistent with a recent investigation of the long time behavior
of the {\em survival} probability.
The localization properties of the eigenstates are
found to be very different from diffusion with no traps.
\end{abstract}
\vspace{5mm}
%
%
%
%
%
\narrowtext
\section{Introduction}
\label{sec:intro}
Diffusion in the presence of randomly distributed traps
appears in many physical situations \cite{HK,HB}.
Well-known examples include the migration
of optical and magnetic excitations in solids \cite{AK}.
Diffusion with traps differs from the usual {\em kinetic} random walk
in an inhomogeneous environment since each diffusive
trajectory must be weighted the same unlike in the latter case
\cite{M,GNMF,GM,GMN}.  This model, known also as the {\em ideal chain},
has received much less attention than the kinetic random walk.

For example, although the case of uncorrelated or weakly correlated disorder
has been fairly well studied \cite{N,W,DV}, only recently was the case
of strongly correlated disorder addressed \cite{GNMF,GM,GMN}.
Unlike in the former case, no rigorous results are known
for many key quantities in the case of the strongly correlated disorder.
For example, the {\em survival} probability, which is the probability
for a walk not to be absorbed by a trap after $t$-steps is not known
rigorously. A recent investigation \cite{GM}, however, showed that it is
possible to extract the asymptotic behavior of the survival probability
for the strongly correlated disorder by considering the full
probability distribution for the number of surviving
walks at the (discrete) time $t$. The result was found to be qualitatively
similar to the case of uncorrelated disorder, the so-called
Donsker-Varadhan behavior \cite{DV}, but quantitatively different
with different numerical values of the exponents.
Specifically an asymptotic, stretched exponential decay of
the probability was predicted where the exponent was related
to the free energy fluctuation exponent $\chi$.

In this work, we first apply the same method as in \cite{GM} to a
closely related but slightly different quantity
(called the {\em survival-return} probability, see below),
which results in a similar prediction of
a stretched exponential tail. This quantity is, however, directly
related to the spectral density of the eigenvalues of the
transition probability matrix which describes the diffusion process.
This allows us to test the prediction by numerically evaluating
the spectral density and comparing with the prediction.
In addition, the behavior of the density of states is itself
of physical interest since it is in principle experimentally measurable,
e.g., by Raman scattering \cite{fontana}.

As a prototype of a system with strongly correlated substitutional
disorder, the incipient infinite percolation cluster \cite{C} in two
(square lattice) and three dimensions (simple cubic lattice) will be
considered.  In this model, perfectly absorbing traps are distributed
with probability $1-p_c$, where $p_c$ is the critical threshold for
the probability for a site to be present in the lattice.
The vacancies thus play the role of the traps.

The work is organized as follows. In Sec.\ \ref{sec:binary} the analogy
to the Schr\"{o}dinger equation is carried out. In Sec.\ \ref{sec:transfer}
the survival-return probability is studied. Sec.\ \ref{sec:lifshitz}
establishes the connection with the density of states which is then
studied numerically in Sec.\ \ref{sec:results}. In Sec.\ \ref{sec:localization}
the localization properties of the eigenmodes are briefly studied and finally
Sec.\ \ref{sec:conclusions} summarizes the conclusions.

\section{Analogy to the Schr\"{o}dinger equation}
\label{sec:binary}
When the total probability is a conserved function of time,
the probability $P_{\0,\x}(t)$ that a random walker is
at the position $\x$ at the (discrete) time $t$ in a $d$ dimensional
hypercubic lattice satisfies the general master equation:
\beqar \label{master_gen}
P_{\0,\x}(t+1) &=& P_{\0,\x}(t) +
    \sum_{\y(\x)} [w_{\x,\y} P_{\0,\y}(t) - w_{\y,\x} P_{\0,\x}(t)]
\eeqar
where $\x \in {\fancy Z}^d$ is spanned by an orthonormal basis
$\{\evm \}_{\mu=1,...,d}$.
For the present problem with traps, however, we assume that
the probability at a trap site is immediately lost
and that the probability at the next time step is determined
solely by the incoming probability from the available neighbor sites.
Therefore, denoting the set of available sites by ${\cal C}$ and
that of traps by ${\cal C}^{\perp}$, we get
\beqar
P_{\0,\x}(t+1) &=& \frac{1}{z} \sum_{\y(\x) \in \clust} P_{\0,\y}(t) ,
 \;\;\;\;\; \mbox{for } \x \in {\cal C} \label{master_chain_2} \\
P_{\0,\x}(t) &=& 0 ,\;\;\;\;\;\; \mbox{for } \x \in {\cal C}^{\perp}
\label{master_chain_3}
\eeqar
where $\y (\x)$ denotes the nearest neighbors of $\x$,
and $z$ is the coordination number of the (full) lattice.

Using the discrete time derivative defined as
\beqar \label{time_derivative}
\frac{\partial}{\partial t} P_{\0,\x}(t) &\equiv&  P_{\0,\x}(t+1)-
P_{\0,\x}(t)
\eeqar
we can write Eq.\ (\ref{master_chain_2}) as:
\beqar \label{master_partial_cont}
\frac{\partial}{\partial t} P_{\0,\x}(t) &=& \frac{1}{z} \sum_{\y}
    [\delta_{1,|\x-\y|} - z \delta_{\y,\x}] P_{\0,\y}(t)
      - \frac{1}{z} V_{\x} P_{\0,\x}(t)
\eeqar
where we defined a hard core potential
\beqar \label{potential_partial}
V_{\x} = \left \{ \begin{array}{lll}
                + \infty ,\;\; &\x \in {\cal C}^{\perp}\;\;&(\mbox{trap}) \\
                  0 ,  &\x \in {\cal C}\;\;&(\mbox{available site})
                  \end{array}
         \right.
\eeqar
Then we can write Eq.\ (\ref{master_partial_cont}) as
\beqar \label{chain_continuous}
\frac{\partial}{\partial t} P_{\0,\x}(t) &=& - H P_{\0,\x}(t)
\eeqar
where we defined the Hamiltonian
\beqar \label{hamiltonian_chain}
H &=& \frac{1}{z} (-\nabla_{\x}^2 + V_{\x})
\eeqar
The discrete spatial derivative is defined as usual as
\beqar \label{spatial_derivative}
\nabla_{\mu} P_{\0,\x}(t)&=& P_{\0,\x+\evm}(t) -P_{\0,\x}(t) .
\eeqar

By analogy to the case of the Schr\"{o}dinger equation for a binary
random alloy under the tight binding approximation \cite{ziman}, one then
expects a stretched-exponential tail (called Lifshitz tail) for the low energy
behavior of the density of states.
The crucial feature for this behavior is the fact that our binary
potential represents a random, infinite barrier which isolates the
low energy band and allows the Lifshitz argument to apply.
This corresponds to the fact that the traps do not allow a stationary
state (frequency $\omega =0$).  In contrast,
the case of the kinetic random walk with conserved probability
cannot be cast in a form of Eq.\ (\ref{chain_continuous}); e.g., for
the myopic ant, there is a dispersion in the first term inside the brackets
on the right-hand side of Eq.\ (\ref{master_partial_cont})
and for the blind ant there is one for the second term in the brackets.
This corresponds to the fact that the conservation
of probability allows a stationary state to exist as well as other states
leading up to it.  We believe that this is why the density of states
in the kinetic random walk problem does not have
a Lifshitz tail \cite{new};
rather, it has a power law behavior in the low-energy limit governed by
the {\em spectral dimension} \cite{AO}.

\section{Survival-return Probability}
\label{sec:transfer}
The master equation (\ref{master_chain_2}) can be put in a transfer matrix
form and numerically solved by the iteration of this matrix \cite{GNMF,GM}.
All the quantities of interest
can then be computed rather accurately, with the exception of the power
moments of multiplicative random variables such as $C\equiv C(\0,t)$,
which is the number of $t$-step walks having the
common starting point $\0$.  This is due to the fact that these moments are
not self-averaging, which forces us to compute
the full probability distribution $P(C,t)$ \cite{GM}.

Similar feature of the lack of self-averaging holds true for the quantity
$C_0 \equiv C_{\0,\0}(t)$ which is the number of walks that return to
the starting point $\0$ after $t$ steps. This quantity determines the
{\em survival-return} probability $P_S^0(t)$ by
\beqar \label{survival-return}
P_S^0(t)=C_{\0,\0}(t)/z^t .
\eeqar

In Fig.\ \ref{fig1}, our numerical result for the distribution
of $\ln C_{\0,\x}(t)$, denoted by $P(\ln C_0)=P(C_0,t) C_{\0,\0}(t)$,
is shown for the square lattice in $d=2$ for various $t$, while
in Fig.\ \ref{fig2} the analogous result for the simple cubic lattice
in $d=3$ is shown. In the numerical work, each disorder configuration was
generated relative to a seed site, which also served as the starting
point $\0$ of the chains of 1600 steps to be exactly enumerated.
The final results were obtained by averaging
over a large number (typically 6000) of independent disorder configurations.

The distribution can be approximated very accurately by a Gaussian;
that is, we may express $P(C_0,t)$ by a log-normal distribution of the form
\beqar \label{log-normal}
P(C_0,t)&=&\frac{1}{C_0\sqrt{2 \pi \sigma_t^2}} \exp[\frac{-(\ln C_0
-\lambda_t)^2} {2 \sigma_t^2}]
\eeqar
where
\beqml
\beqar
\lambda_t &=& \overline{\ln C_{\0,\0}(t)} \label{mean_var:1} \\
\sigma_t^2 &=& \overline{(\ln C_{\0,\0}(t))^2} \label{mean_var:2}
-\overline{\ln C_{\0,\0}(t)}^2
\eeqar
\eeqml
The mean $\lambda_t$ and the variance $\sigma_t^2$ both depend on
the time $t$. This dependence can be deduced from the log-normal fit
but can also be directly evaluated from the logarithmic moments of $C_0$
which are much easier to obtain numerically than the power moments.
The long-time behavior for these two quantities can be fit rather well by
\beqml
\beqar \label{asymptote}
\lambda_t &\sim& t \ln z_{eff} -\alpha t^{\psi_0} \label{mean_var2:1} \\
\sigma_t^2 &\sim& \beta t^{2 \chi_0} \label{mean_var2:2}
\eeqar
\eeqml
where $\alpha$, $\beta$, and $z_{eff}$ are contants.

This behavior is similar to the case of the {\em survival} probability
\beqar \label{survival}
P_S(t)=C(\0,t)/z^t
\eeqar
as discussed in Ref.\cite{GM}.

The best fit values of the exponents are summarized in Table\ \ref{table1}.
The calculations based on the log-normal fit and those directly from
the logarithmic moments give consistent results as shown.
In particular the value of the {\em effective} coordination number $z_{eff}$
in $d=2$ is $z_{eff}=3.62 \pm 0.01$ (from log-normal) and
$z_{eff}=3.60 \pm 0.01$ (from direct calculation).
In $d=3$, we find $z_{eff}=4.33 \pm 0.14$ and $z_{eff}=4.12 \pm 0.11$ from
the two procedures, respectively.
The higher relative value $z_{eff}/z$ in $d=2$ than $d=3$ is not
surprising and is due to the much more compact structure of the
incipient infinite cluster in two dimensions.
The effect of the traps is indeed to force the
diffusion into regions of high connectivity in order to maximize
the entropy \cite{GM,GMN}.

The fact that $\sigma_t^2 \gg \lambda_t$ for sufficiently long times has a
consequence that the log-normal distribution has to be cut off both at
the short and long times as discussed in \cite{GM} for the case
of the survival probability. If this were not the case
all the moments of $C_{\0,\0}(t)$, which can be easily computed
from Eq.\ (\ref{log-normal}), would grow faster than $z^t$, which is clearly
impossible.  Therefore, we assume a cutoff $1 \leq C_{\0,\0}(t) \leq z^t$
(which is certainly true), and carry out the calculation of the moments
in the same manner as shown in Ref.\cite{GM,GMN} for $C(\0,t)$.
This calculation yields a Donsker-Varadhan type behavior \cite{DV}
for the first moment of $C_{\0,\0}(t)/z^t$,
\beqar \label{final_0}
\overline{P_S^0(t)} &\sim& e^{-t^{2(1-\chi_0)}}
\eeqar
where $\chi_0$ is the exponent appearing in Eq.\ (\ref{asymptote}).
This is qualitatively similar to the behavior of the {\em survival}
probability $P_S(t)$ \cite{GM,GMN,note} (but generally with different numerical
value for the exponent).  However, it is even qualitatively distinct
from the behavior of the {\em return} probability
\beqar \label{return}
P_0(t)=C_{\0,\0}(t)/C(\0,t) ,
\eeqar
which was also calculated in Ref.\cite{GM} and found to be a power law in $t$
for large $t$.

\section{Density of States}
\label{sec:lifshitz}
The transition probability matrix ${\bf W}$ is a random, non-negative definite
matrix of size $S\times S$ where $S({\clust})$ is the number of available
sites of a particular configuration ${\clust}$ and has elements $W_{\x,\y}$
given by
\beqar \label{transition}
W_{\x,\y} &=& \left \{ \begin{array}{ll}
                1/z \;\; &\mbox{if $|\x-\y|=1$ and $\x,\y \in \clust$}\\
                0                &\mbox{otherwise}
                \end{array}
                \right.
\eeqar
We will connect the spectral properties of this matrix to the behavior of
the survival-return probability $P_S^0(t)$ in the same way as for the
kinetic random walks \cite{MNF}.  This relation will be completely general
and valid in any dimensions and any amount of disorder.

The uniformly weighted mean of $P_S^0(t)$ over all starting points of a
given disorder configuration of size $S$ is simply
\beqar
\langle P_S^0(t) \rangle = \sum_{\0 \in \clust} C_{\0,\0}(t)/(z^t S)
\eeqar
For a given starting point, we have
\beqar
C_{\0,\0}(t)/z^t = {\bf e}_{\0}^{T} {\bf W}^t {\bf e}_{\0} ,
\eeqar
where ${\bf e}_{\0}$ is the column vector whose components are all zeros
except the component (which is one) corresponding to the site $\0$ and
$ {\bf e}_{\0}^{T} $ is the corresponding row vector of $ {\bf e}_{\0} $.
The superscript $T$ refers to the transpose in the matrix notation.
Thus,
\beqar
\label{eq:trace}
\langle P_S^0 (t)\rangle = \mbox{Tr} {\bf W}^t / S
= \sum_{i}{\lambda_{i}}^t /S
\eeqar
where $\lambda_i$ are the eigenvalues of ${\bf W}$.

For the case of bipartite medium (e.g., a cluster on the square lattice or
the simple cubic lattice), the eigenspectrum is symmetric about zero.
In this case, $\langle P_S^0 (t) \rangle$ is zero for all odd time steps
$t$ and twice the sum only over the positive eigenvalues for the even
times $t$.  (For non-bipartite cases this is no longer true, but the
asymptotic long time behavior should be largely determined by the part of
the spectrum near the maximum in any case.)  For the bipartite cases we have
\beqar \label{survival_return_2}
\langle P_S^0(t)\rangle = (2/S) \sum_{i} e^{-\epsilon_{i}t}
\eeqar
where the sum is over the positive spectrum and we let the eigenvalues
$\lambda_{i}=e^{-\epsilon_{i}}$ with $\epsilon_{i} > 0$.
Laplace transforming this, we get
\beqar \label{Laplace_transform}
\Ptilde_S^0(\omega) &=& {\cal L} [\langle P_S^0(t)\rangle] =
     (2/S) \sum_{i} \int_0^{\infty} dt e^{-(\omega + \epsilon_{i})}
       = (2/S) \sum_{i} \frac{1}{\omega + \epsilon_{i}}
\eeqar
Then using the well known identity:
\beqar \label{identity2}
\frac{1}{x+i 0^+}&=& P(\frac{1}{x}) - i \pi \delta(x)
\eeqar
where $P$ stands for the principal part in the sense of the distributions,
we obtain
\beqar \label{final_result}
\rho(\epsilon)= - \frac{1}{\pi} \mbox{Im}
\overline{\Ptilde_S^0(-\epsilon + i 0^+)}
\eeqar
where the disorder-averaged density of states $\rho (\epsilon )$
was defined to be
\beqar \label{density_states}
\rho(\epsilon) =  \overline{(2/S) \sum_{i} \delta(\epsilon-\epsilon_{i})}
\eeqar

If we now assume a stretched exponential behavior for the survival-return
probability in the long time limit as discussed earlier,
\beqar \label{saddle_point_result}
\overline{P_S^0(t)} \sim e^{-t^{\frac{\tilde{d_0}}{\tilde{d_0}+2}}}
\eeqar
then Eq.\ (\ref{final_result}) predicts for $\epsilon \rightarrow 0$,
\beqar \label{dtilde_0_def}
\rho(\epsilon) \sim e^{-\epsilon^{-\tilde{d_0}/2}}
\eeqar
which is the corresponding Lifshitz tail for the density of states
\cite{note1}.

The exponent $\tilde{d_0}$, defined by Eq.\ (\ref{dtilde_0_def}),
can be thought of as the analog of the spectral dimension \cite{AO}
of the usual kinetic random walk. It should be stressed, however that,
unlike that case, $\tilde{d_0}$ is not a power-law exponent and is not
related to the {\em return} probability but to the {\em survival-return}
probability.

\section{Numerical results}
\label{sec:results}
In order to numerically evaluate the density of states $\rho (\epsilon )$,
we need to diagonalize a large number of large matrices since
the thermodynamic limit of a large system is needed and
a disorder average over many such systems must be taken.
This would ordinarily present a challenge; however,
a numerical method using the algorithm developed by Saad \cite{S}
conveniently allows us to approximately diagonalize the matrices and
obtain the eigenspectrum near the extrema with high accuracy.
This method was already successfully exploited
in the context ot the kinetic random walks (the so-called {\em ants})
\cite{MNF}.

Since the matrix ${\bf W}$ is symmetric, all the eigenvalues are real and,
moreover, they are contained in the interval $(-1,1)$.
The maximum eigenvalue $\lambda_{max}$, which has multiplicity
one as ensured by the Perron-Frobenious theorem \cite{P}
for non-negative matrices,
can be interpreted as the ratio $z_{eff}/z$ of the effective
coordination number to the full coordination number of the lattice.
It should be noted, however, that ${\bf W}$ is {\em not} a
Markov matrix, due to the non-conservation of the probability
\cite{note2}.

The disordered media were created as critical percolation clusters,
starting from a seed site and growing layer by layer in a breadth-first
fashion and stopping growth when a predetermined number of sites were
generated. In both $d=2$ and $3$, we used clusters with a fixed size of
$10000$, requiring ${\bf W}$ of $10000 \times 10000$.
The exponent $\tilde{d_0}$ have been calculated from the numerically
obtained density of states using Eq.\ (\ref{dtilde_0_def})
and the results are shown in Fig.\ \ref{fig3}.
We employed $2500$ configurations to
extract the values of $\tilde{d_0}$ in both $two$ and $three$ dimensions.
{}From Eqs.\ (\ref{final_0}) and (\ref{saddle_point_result}), the relation
\beqar \label{scaling_chi0}
\frac{\tilde{d_0}}{2}&=& \frac{2(1-\chi_0)}{2 \chi_0 -1}
\eeqar
follows, allowing the evaluation of $\chi_0$ from that of $\tilde{d_0}$.
The final exponent estimates from the density of states are summarized
in Table\ \ref{table2}, where we have also repeated the estimates of
$\chi_0$ from Table\ \ref{table1}.

As shown in Fig.\ \ref{fig3}, the qualitative agreement with
Eq.\ \ref{dtilde_0_def}) is very clear for both $d=2$ and $3$.
Quantitatively, the spectral and exact enumeration estimates of
$\chi_0$ seem to be in excellent agreement for $d=2$,
while those for $d=3$ do not appear to agree as well.
In particular the trend with the dimensionality seems to be reversed
depending on the method of evaluation.
Also the value of $z_{eff}$ from the maximum eigenvalue $\lambda_{max}$
turns out to be $3.48 \pm 0.04$ and $3.67 \pm 0.04$ in $d=2$ and $d=3$,
respectively.  When compared with the exact enumeration estimates as
discussed in Sec.\ \ref{sec:transfer},
it is apparent that the spectral calculation in $d=2$ is much more consistent
with the exact enumeration result than in $d=3$.

The cause of this quantitative discrepancy is still unclear and it will
require further work to determine whether it is simply a finite size
effect or indicative of a deeper difference between two- and three-dimensional
systems, perhaps due to the much less compact structure of the disordered media
in three dimensions.

\section{Localization properties}
\label{sec:localization}
It is well-known that, in general, the localization properties
of the eigenstates of a given Hamiltonian are strongly
affected by the presence of quenched disorder \cite{Lifshitz}.
Less efforts have been devoted, however, to the cases where the
disorder is critical, as in the present problem.

We shall present here some preliminary
results, based on the numerical investigations of the spectrum
in the case of small matrices, but which nevertheless allow us
to obtain a qualitative idea of the difference with the case in the
absence of traps.
A more detailed investigation along with a careful comparison
with the case of the kinetic random walk are still in progress
and will be published subsequently.

In Fig.\ \ref{fig4} we show the entire spectra averaged over a number of
typical configurations in $d=2$ and $3$.
The symmetry about $\lambda =0$ is due to the bipartite nature of the
lattices used, and the absence of the eigenvalue with magnitude 1
(or close to it) reflects the lack of the conservation of probability.
Clearly there is no trace of the accumulation of modes near the maximum
as is the case for the {\em ants} \cite{MNF}.

A quantitative measurement of the degree of localization of a mode
is given by the {\em participation ratio} \cite{T}:
\beqar \label{participation}
P(\lambda) &=& \frac{\sum_{\x} |u_{\x}(\lambda)|^4}
{(\sum_{\x}|u_{\x}(\lambda)|^2)^2}
\eeqar
where $u_{\x}$ is the amplitude at $\x$ of the mode.
With this definition, we have $P(\lambda) \in [0,1]$ and a value close
to $1$ corresponds to a well localized state, while a small value
(of order of $1/S$ where $S$ is the number of sites in the cluster
${\cal C}$) corresponds to an extended state.

We diagonalized exactly a small matrix of $400 \times 400$ for both $d=2$
and $d=3$. The results for $P(\lambda)$ are presented in
Fig.\ \ref{fig5},\ref{fig6} in $d=2$ and $3$, respectively.
Even though all states are localized,
it appears that some modes are particularly strongly localized,
such as those corresponding to the extreme negative part of the
spectrum, while some are more closely resembling an extended state.
To illustrate this, we plot in Fig.\ \ref{fig7} the amplitudes of some
chosen modes in the $d=2$ case. Fig.\ \ref{fig7}{\sl a} and {\sl b} correspond
to the two extremal eigenvalues (negative and positive) and {\sl c} to
a less localized state at $\lambda=0$.
These structures seem to be consistent with their participation ratios.
Similar results persist also for different configurations.

\section{Conclusions}
\label{sec:conclusions}
The main aim of the present work is to confirm the existence of
a stretched exponential tail in the long time behavior of the
various probabilities for diffusion with critically disordered traps.
This has been accomplished by comparing the predicted values of the
appropriate exponents based on the exact enumeration results
with the corresponding Lifshitz tail in the density of states of
the transition probability matrix.

Numerical results indicate that the two procedures give
{\em qualitatively} consistent results in both two and three
dimesions.  {\em Quantitative} agreement is also excellent in two
dimensions; however, it is substantially poorer in three dimensions.
More work will be necessary to determine the cause of this
numerical discrepancy.

The localization properties of the eigenstates of this system
were also investigated using mainly the participation ratio.
It was found that, although all states seem to have a localized
character, a wide range of the degree of localization, ranging
from almost extended to fully localized, are present.
In particular the lowest extreme of the spectrum (in the negative
part, which does not seem to play any role in the
determination of the long time diffusional behavior) is extremely localized.

\acknowledgments
Enlightening discussions with Norm Fuchs, Don Jacobs, Yong-Jihn Kim
and Sonali Mukherjee are gratefully acknowledged.

%
%

%
%
\begin{figure}
\caption{Calculated distribution $P(\ln C_0)$, for $t=400 \; (\bigcirc)$,
$t=800 \; (\bigtriangleup)$, $t=1200 \; (+)$, $t=1600 \; (\Diamond)$,
for the square lattice in $d=2$. The solid curves are the best fit results
derived from Eq.\ (\protect{\ref{log-normal})}. The values of
$C_0$ were normalized by an arbitrary factor $2.8$ for convenience. }
\label{fig1}
\end{figure}
\begin{figure}
\caption{Calculated distribution $P(\ln C_0)$, for $t=400 \; (\bigcirc)$,
$t=800 \; (\bigtriangleup)$, $t=1200 \; (+)$, $t=1600 \; (\Diamond)$,
for the simple cubic lattice in $d=3$. The solid curves are the best fit
results
derived from Eq.\ (\protect{\ref{log-normal})}. The values of
$C_0$ were normalized by an arbitrary factor $3.0$ for convenience. }
\label{fig2}
\end{figure}
\begin{figure}
\caption{Evaluation of the exponent $\tilde{d_0}/2$ for $d=2$
$(\bigcirc)$ and $d=3$ $(\bigtriangleup)$.
The solid lines are the best fit results yielding
$\tilde{d_0}/2=2.69 \pm 0.04$ and $4.42 \pm 0.14$ in $d=2$ and $3$,
respectively. The corresponding $\chi_0$ is given in
Table\ \protect{\ref{table2}}.}
\label{fig3}
\end{figure}
\begin{figure}
\caption{Eigenspectra of a typical $400 \times 400$ transition probability
matrix in $d=2$ ($\bigcirc$) and $d=3$ (+) at critical disorder.}
\label{fig4}
\end{figure}
\begin{figure}
\caption{Participation ratio $P(\lambda)$ for the spectrum of a
$400 \times 400$ transition probability matrix in $d=2$ at critical disorder.}
\label{fig5}
\end{figure}
\begin{figure}
\caption{Participation ratio $P(\lambda)$ for the spectrum of a
$400 \times 400$ transition probability matrix in $d=3$ at critical disorder.}
\label{fig6}
\end{figure}
\begin{figure}
\caption{Plot of the ${\bf x}$ dependence of the amplitude
$u_{\x}(\lambda)$ of the eigenmodes corresponding to the lower
($\lambda=-0.8390..$), upper ($\lambda=0.8390..$) and middle ($\lambda=0$)
part of the spectrum (in {\sl a,b,c} respectively) for the two-dimensional
case.}
\label{fig7}
\end{figure}
%
%
%
\begin{table}
\caption{Summary of the exponents $\psi_0$ and $\chi_0$ defined in the text.
The label (D) and (I) mean direct evaluation and from the
log-normal distribution, respectively. }
\begin{tabular}{lcccc}
\multicolumn{1}{l}{$d$}&
\multicolumn{1}{c}{$\psi_0$(D)}&
\multicolumn{1}{c}{$\psi_0$(I)}&
\multicolumn{1}{c}{$\chi_0$(D)}&
\multicolumn{1}{c}{$\chi_0$(I)}\\
\hline
$2$&$0.70 \pm 0.01$&$0.69 \pm 0.01$
&$ 0.65 \pm 0.01$&$0.65 \pm 0.01$ \\
$3$&$0.78 \pm 0.03$ &$0.83 \pm 0.03$
&$ 0.72 \pm 0.01$&$0.70 \pm 0.01$ \\
\end{tabular}
\label{table1}
\end{table}
\begin{table}
\caption{Estimates of the exponents $\tilde{d_0}$ and $\chi_0$
from the density of states analysis.  For comparison, those of
$\chi_0$ calculated from the survival-return probability are also
repeated from Table\ \protect{\ref{table1}}.}
\begin{tabular}{lcccc}
\multicolumn{1}{l}{$d$}&
\multicolumn{1}{c}{$\tilde{d_0}$}&
\multicolumn{1}{c}{$\chi_0$} &
\multicolumn{1}{c}{$\chi_0$(D)$^a$}&
\multicolumn{1}{c}{$\chi_0$(I)$^a$}\\
\hline
$2$& $5.38 \pm 0.08$ &$0.64 \pm 0.01$&$0.65 \pm 0.01$&$0.65 \pm 0.01$ \\
$3$& $8.84 \pm 0.28$ &$0.59 \pm 0.01$&$0.72 \pm 0.01$&$0.70 \pm 0.01$ \\
\end{tabular}
$^a$ From Table\ \protect{\ref{table1}}.
\label{table2}
\end{table}

\begin{references}
\bibitem{note}As pointed out in \cite{GM}, the analysis of the power moments
of $C$ given in Ref.\cite{GNMF} appears to be incorrect, having been based on
on an insufficient numerical evidence.
\end{references}
\end{document}